\RequirePackage{rotating}
\documentclass[fleqn,usenatbib]{mnras}

\usepackage{newtxtext,newtxmath}
 
\usepackage[T1]{fontenc}

\DeclareRobustCommand{\VAN}[3]{#2}
\let\VANthebibliography\thebibliography
\def\thebibliography{\DeclareRobustCommand{\VAN}[3]{##3}\VANthebibliography}

\usepackage{graphicx}	
\usepackage{amsmath}	
\usepackage{amssymb}    
\usepackage{amsfonts}
\usepackage{appendix}
\usepackage{rotating}
\usepackage{pdflscape}
\usepackage{longtable}
\usepackage{comment}

\title[Constraining the Neutral Fraction]{Inferring the IGM Neutral Fraction at z $\sim$ 6-8 with Low-Luminosity Lyman Break Galaxies}

\author[Bolan et al.]{Patricia Bolan,$^{1}$
Brian C. Lemaux,$^{1,2}$
Charlotte Mason,$^{3,4,5,\dagger}$
Maru$\Breve{\textrm{s}}$a Brada$\Breve{\textrm{c}}$, $^{1}$
Tommaso Treu,$^{6}$
\newauthor
Victoria Strait,$^{1,4}$
Debora Pelliccia,$^{7}$
Laura Pentericci,$^{8}$
Matthew Malkan$^{6}$
\\
$^{1}$Department of Physics and Astronomy, University of California, Davis, 1 Shields Ave, Davis, CA 95616, USA\\
$^{2}$Gemini Observatory, NSF’s NOIRLab, 670 N. A’ohoku Place, Hilo, Hawai’i, 96720, USA\\
$^{3}$Center for Astrophysics $|$ Harvard \& Smithsonian, 60 Garden St., Cambridge, MA 02138, USA\\
$^{4}$Cosmic Dawn Center (DAWN)\\
$^{5}$Niels Bohr Institute, University of Copenhagen, Jagtvej 128, København N, DK-2200, Denmark\\
$^{6}$Department of Physics and Astronomy, University of California, Los Angeles, CA 90095-1547, USA\\
$^{7}$UCO/Lick Observatory, Department of Astronomy $\&$ Astrophysics, UC Santa Cruz, 1156 High Street, Santa Cruz, CA 95064, USA\\
$^{8}$INAF Osservatorio Astronomico di Roma, Via Frascati 33, I-00040 Monteporzio (RM), Italy\\
$^{\dagger}$Hubble Fellow 
}

\date{Accepted XXX. Received YYY; in original form ZZZ}

\pubyear{2021}

\begin{document}
\label{firstpage}
\pagerange{\pageref{firstpage}--\pageref{lastpage}}
\maketitle

\begin{abstract}
We present a Bayesian inference on the neutral hydrogen fraction of the intergalactic medium (IGM), $\overline{x}_\textsc{hi}$, at $z \sim$ 6-8 using the properties of Lyman break galaxies during the Epoch of Reionization. We use large samples of LBG candidates at $5.5 \leq z \leq 8.2$ with spectroscopy from Keck/DEIMOS and Keck/MOSFIRE. For each galaxy, we incorporate either the Lyman-$\alpha$ equivalent width (EW) for detections or the EW limit spectrum for nondetections to parameterize the EW distribution at various ultraviolet brightnesses for a given redshift. Using our reference sample of galaxy candidates from the ionized universe at $z$ $\sim$ 6.0, we are able to infer $\overline{x}_\textsc{hi}$ at two redshifts: $z$ $\sim$ 6.7 and $z$ $\sim$ 7.6. This work includes intrinsically faint, gravitationally lensed galaxies at $z$ $\sim$ 6.0 in order to constrain the intrinsic faint-end Ly$\alpha$ EW distribution and provide a comparable population of galaxies to counterparts in our sample that are at higher redshift. The inclusion of faint galaxy candidates, in addition to a more sophisticated modelling framework, allows us to better isolate effects of the interstellar medium and circumgalactic medium on the observed Lyman-$\alpha$ distribution from those of the IGM. We infer an upper limit of $\overline{x}_\textsc{hi}$ $\leq$ 0.25 at $z$ = 6.7 $\pm$ 0.2 and a neutral fraction of $\overline{x}_\textsc{hi}$ = $0.83^{+0.08}_{-0.11}$ at $z$ = 7.6$\pm$ 0.6, both within 1$\sigma$ uncertainty, results which favor a moderately late and rapid reionization.

\end{abstract}

\begin{keywords}
galaxies -- reionization -- high-redshift galaxies
\end{keywords}



\section{Introduction}

When the Universe was less than one billion years old, neutral hydrogen atoms in the intergalactic medium (IGM) were ionized by the first light sources. This period of time is called Epoch of Reionization (EoR) and is thought to lie in the redshifts $z \sim 6-10$ \citep{fan_observational_2006, schroeder_evidence_2013, hinshaw_nine-year_2013, mcgreer_model-independent_2015}. Some of the biggest open questions in astronomy concern a detailed timeline of the EoR and the sources which are responsible for it, neither of which have been entirely addressed \citep{robertson_cosmic_2015, robertson_galaxy_2021}. The first light sources in the Universe were young, star-forming galaxies and quasars. The steep faint end of the UV luminosity function (LF) of galaxies suggests that faint galaxies exist in large quantities beyond the observational detection threshold of most surveys of galaxies in this epoch \citep{bouwens_reionization_2015, bouwens_z_2017, finkelstein_evolution_2015, livermore_directly_2017}. Intrinsically faint galaxies are likely promising candidates as the drivers of reionization due to their large numbers if they are capable of producing even modest amounts of escaping ionizing photons \citep[e.g.][]{marchi_ly_2018, steidel_keck_2018, izotov_low-redshift_2018, finkelstein_conditions_2019, pahl_uncontaminated_2021}; however, there is also the possibility of significant ionization from bright galaxies \citep{naidu_rapid_2020, jung_texas_2020}. 

One powerful probe of the EoR is Lyman-alpha emission (Ly$\alpha$, 1216\AA), as it is intrinsically the strongest line in the UV. Ly$\alpha$ photons are attenuated by neutral hydrogen, making the line a probe of the ionization state of the IGM as well as properties of the sources which emitted them \citep{haiman_models_1999, verhamme_3d_2006, mcquinn_studying_2007, dijkstra_ly_2014}. Searching for Ly$\alpha$ emission from samples of galaxies with for multiple lines of sight provides a probe of the state of the IGM throughout reionization, provided that the intrinsic emission of Ly$\alpha$ is known. Thus, Ly$\alpha$ emission allows us to trace the volume-averaged neutral hydrogen fraction, $\overline{x}_\textsc{hi}$, of the IGM, which starts at $\overline{x}_\textsc{hi}$ = 1 at the onset of reionization, and ends at $\simeq$ 0 by $z$ $\sim$ 6 \citep{fan_constraining_2006, miralda-escude_reionization_1998}. To achieve a more detailed timeline of reionization, it is necessary to determine the evolution of $\overline{x}_\textsc{hi}$ throughout the EoR. 

A common method used to determine $\overline{x}_\textsc{hi}$ at a given redshift is to calculate the Ly$\alpha$ emitter fraction: a sample of photometrically-selected galaxies is obtained via the Lyman break technique, and the number of them which emit Ly$\alpha$ photons above an equivalent width (EW) threshold is determined in a spectroscopic follow up. This method has yielded results which reflect an increase in Ly$\alpha$ optical depth and IGM neutral fraction from $z$ = 6 to $z$ = 7 \citep[e.g][]{kashikawa_end_2006, kashikawa_completing_2011, pentericci_spectroscopic_2011, pentericci_new_2014, fontana_lack_2010, schenker_keck_2012, schenker_line-emitting_2014, tilvi_rapid_2014, ono_spectroscopic_2012, treu_inferences_2012, fuller_spectroscopically_2020, jung_texas_2020}. These methods have been instrumental in constraining the reionization timeline, but it is difficult to determine a $\overline{x}_\textsc{hi}$ value from a simple Ly$\alpha$ emitter fraction. Before photons can travel through the IGM, they must escape the interstellar medium (ISM) and circumgalactic medium (CGM) of a galaxy, which affects Ly$\alpha$ photons, as they get scattered by neutral hydrogen in the ISM and CGM before escaping the galaxy. In addition, the large-scale structure of the IGM may have effects on the neutral fraction inference, due to the likely patchiness of neutral hydrogen clouds \citep{trac_imprint_2008, becker_evidence_2018, daloisio_heating_2019}. In the method employed in this paper, introduced by \citet{mason_universe_2018} and \citet{mason_inferences_2019}, effects of the ISM and CGM are isolated from those of the IGM by attributing any changes in Ly$\alpha$ emission in the reionizing universe to partial IGM opacity. In addition, we use simulations with realistic distributions of neutral hydrogen to investigate the effects of patchiness, providing a path to measuring $\overline{x}_\textsc{hi}$. 

We therefore use the full Ly$\alpha$ EW distribution from samples of Lyman Break Galaxies (LBGs) to infer the IGM neutral fraction. By incorporating information from the entire sample, both galaxies with detected Ly$\alpha$ emission and those without, the distribution of the Ly$\alpha$ optical depth and the neutral fraction can be constrained to high precision. We use a Bayesian method to infer $\overline{x}_\textsc{hi}$ using a sample of galaxies at various redshifts during the EoR. In this work, we incorporate a new reference sample at $z$ $\sim$ 6.0, when reionization is thought to be largely complete, in order to model Ly$\alpha$ EW after its escape from the ISM, but before encountering the IGM. We require galaxies to have similar inherent properties.

An expansive sample of lensed $z$ $\sim 5-7$ galaxies was recently compiled by \citet{fuller_spectroscopically_2020}, containing spectroscopic data of 198 LBG candidates, 36 of them with Ly$\alpha$ detected in emission. This is the largest faint ($L \sim 0.1L^*$ where $L^*$ is the characteristic luminosity defined by \citealp{bouwens_reionization_2015}) sample at this redshift, assembled from hundreds of orbits on the \emph{Hubble Space Telescope} (\emph{HST}), \emph{Spitzer Space Telescope} (\emph{Spitzer}), and an over four-year long campaign on Keck/DEIMOS \citep{faber_deimos_2003}. With over 100 hours of spectroscopy on these LBG candidates, the depth of these data provides excellent constraints on the EW distribution at $5 \leq z \leq 7$. From \citet{fuller_spectroscopically_2020}, we draw two samples: the reference sample at $z$ $\sim$ 6.0 and a set of galaxy candidates at $z$ $\sim$ 6.7. This dataset compliments our comparable sample of galaxies firmly in the EoR at $z \sim 7.6$ compiled by \citet{hoag_constraining_2019}. The three samples used in this work all have similar luminosity distributions comprised of faint luminosities, representing typical galaxies at these redshifts \citep{hoag_constraining_2019}. The reference sample is exemplary for characterizing the EW distribution at $z$ $\sim$ 6.0 for use in our inference, as the absolute magnitudes of these candidates match those of the higher redshift LBG candidates. This method combines reionization IGM simulations from \citet{mesinger_evolution_2016} with empirical models of the effects of the ISM and CGM, determined via the $z$ $\sim$ 6.0 reference sample, to infer the global neutral hydrogen fraction from LBG properties. We perform the Bayesian inference of $\overline{x}_\textsc{hi}$ at $z$ $\sim$ 6.7 and $z$ $\sim$ 7.6 using this faint galaxy sample in our ionized baseline in order to compare similar galaxy populations at both redshifts. 

The paper is organized as follows. In Section 2, we discuss the data and methods used in this analysis. Section 3 describes the data analysis involved in comparing the three galaxy samples and the neutral hydrogen fraction inference. In Section 4, we present and discuss our results, and conclusions can be found in Section 5. The following cosmology is used in data analysis throughout the paper: $\Omega_{m} = 0.3, \Omega_{\Lambda} = 0.7, H_0 = 70$, and values from \citet{planck_collaboration_planck_2016} for all analysis used to infer the neutral fraction. All magnitudes are given in the AB system, and all equivalent widths are presented in the rest frame. 

\section{Data and Methods}

The data used in this analysis come from two sets of observations: a sample of galaxy candidates between $z \sim 5.5-6.5$ and $z \sim 6.5-7$ \citep{fuller_spectroscopically_2020}, and another set between $z$ $\sim 7-8.2$ \citep{hoag_constraining_2019}. The candidates are detected behind 10 massive lensing clusters, providing a wide range of sightlines which help to alleviate cosmic variance. All imaging data comes from \emph{HST} and \emph{Spitzer} and is summarized in Table \ref{table:obs}. Five of the clusters are from Hubble Frontier Fields (HFF, \citealp{lotz_frontier_2017}): A2744, MACS0416, MACS0717, MACS1149, and A370. Four clusters come from the Cluster Lensing and Supernova Survey with Hubble (CLASH, \citealp{postman_cluster_2012}), MACS0744, MACS1423, MACS2129, and RXJ1347, and the last, MACS2214, has \emph{HST} imaging from the Spitzer UltRa Faint SUrvey Program (SURFSUP, \citealp{bradac_spitzer_2014}). In addition to \emph{HST} imaging from these programs, each cluster has \emph{Spitzer} observations from SURFSUP and the \emph{Spitzer} HFF programs.

\begin{table*}Summary of Observations
    \begin{center}
    \begin{tabular}{||c c c c c c||} 
     \hline
    Cluster Name & Short Name & $\alpha_{J2000}$ (deg) & $\delta_{J2000}$ (deg) & $z_{\textrm{cluster}}$ & \emph{HST} Imaging (\emph{Spitzer} Imaging) \\ 
    \hline\hline
    Abell 2744 &  A2744 & 3.5975000 & -30.39056 & 0.308 & HFF\\ 
    Abell 370 & A370 & 39.968000 & -1.576666 & 0.375 & HFF\\
    MACSJ0416.1-2403 & MACS0416 & 64.039167 & -24.06778 & 0.420 & HFF/CLASH\\
    MACSJ0717.5+0745 & MACS0717 & 109.38167 & 37.755000 & 0.548 & HFF/CLASH (SURFSUP) \\
    MACSJ0744.8+3927 & MACS0744 & 116.215833 & 39.459167 & 0.686 & CLASH (SURFSUP)\\
    MACSJ1149.5+2223 & MACS1149 & 177.392917 & 22.395000 & 0.544 & HFF/CLASH (SURFSUP)\\
    MACSJ1423.8+2404 & MACS1423 & 215.951250 & 24.079722 & 0.545 & CLASH (SURFSUP)\\
    MACSJ2129.4-0741 & MACS2129 & 322.359208 & -7.690611 & 0.568 & CLASH (SURFSUP)\\
    MACSJ2214.9-1359 & MACS2214 & 333.739208 & -14.00300 & 0.500 & SURFSUP\\
    RXJ1347.5-1145 & RXJ1347 & 206.87750 & -11.75278 & 0.451 & CLASH (SURFSUP)\\ [1ex] 
     \hline
    \end{tabular}
    \caption{Summary of observations for 10 lensing clusters used in this work. }
    \label{table:obs}
    \end{center}
\end{table*}

The photometric measurements for the \citet{fuller_spectroscopically_2020} $z \sim 5.5-7$ sample are from the ASTRODEEP \citep{castellano_astrodeep_2016, bradac_hubble_2019} team for all HFF clusters. For the remaining five clusters, we use an identical method to that of ASTRODEEP for photometric calculations. LBG candidates were selected via the Lyman break technique, and 198 were followed up spectroscopically with Keck/DEIMOS. Details about the sample, photometry, and spectroscopy of the $z \sim 5.5-7$ candidates can be found in \citet{fuller_spectroscopically_2020}.

The galaxies at $z \sim 7-8.2$ are taken from \citet{hoag_constraining_2019}. Nine of the cluster fields, A2744, A370, MACS0416, MACS0744, MACS1149, MACS1423, MACS2129, MACS2214, and RXJ1347, are used to search for Ly$\alpha$ emission in the redshift range $z \sim 7-8.25$. The photometric data are used to obtain photometric redshift probability distributions, $P(z)$s. For the full $z \sim 7.6$ sample, the photometry and selection criteria (at least 1$\%$ probability of being in the redshift range) are identical to those used for the \citet{fuller_spectroscopically_2020} sample. More information on the observations of these clusters is detailed in \citet{hoag_constraining_2019}. In this sample of 68 LBG candidates with spectroscopic data from the KECK Multi-Object Spectrometer for InfraRed Exploration (MOSFIRE; \citealp{mclean_design_2010}), there are two confident Ly$\alpha$ detections and upper limits on flux and EW for the other galaxies.

\begin{figure*}
\includegraphics[width=18cm]{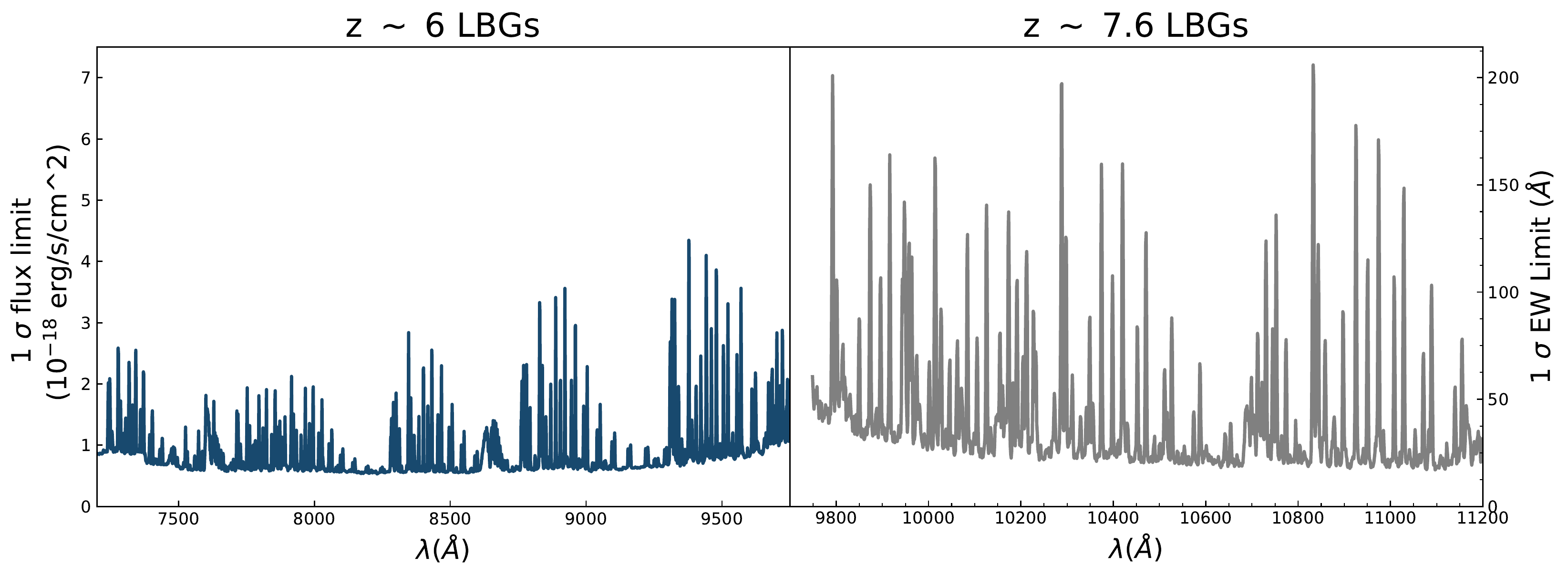}
\caption{Median 1$\sigma$ upper limit on rest-frame Ly$\alpha$ EW and flux as a function of wavelength for DEIMOS (left) and MOSFIRE (right) non-detections at $z$ $\sim$ 6 and 7.6}
\label{fig:limit}%
\end{figure*}

For our analysis, we break up the former galaxy sample into two different redshift bins: $z$ $\sim$ 6.0 and $z$ $\sim$ 6.7 Both samples are drawn from the same set of observations \citep{fuller_spectroscopically_2020}. To select galaxies that will serve as the $z \sim$ 6.0 reference sample, we make a cut based on the fraction of the $P(z)$ within the redshift range of $P(5.5 < z < 6.5)$ for the $z$ $\sim$ 6.0 sample and $P(6.5 < z < 6.9)$ for the $z \sim$ 6.7 sample. Requiring that at least 20$\%$ of the $P(z)$ lies in these redshift ranges yields a sample of 74 reference galaxies at $z \sim 6.0$ and 28 at $z \sim 6.7$. We use $z \sim$ 6.7 as the fiducial redshift for candidates with $P(z > 6.5)$ since this is the median of our Ly$\alpha$ detection window for this sample. For the MOSFIRE sample at $z \sim 7.6$, we use the range $P(7 < z < 8.2)$.  We test whether selecting samples based on $P(z)$ yields different results on our main analysis from selection based on $z_{\textrm{phot}}$, where $z_{\textrm{phot}}$ is the peak value given by the $P(z)$ distribution. We find that there is very little effect as detailed in Section 4.1. While we use these $P(z)$ cutoffs to determine the LBG candidates which will be used in the inferences, the full $P(z)$s of the $z$ $\sim$ 6.7 and $z \sim$ 7.6 galaxies are used as an effective weight of their significance in the neutral fraction inference as described in Section \ref{sec:nfi}.

\subsection{Flux Calibration}
With the three samples of LBG candidates, all with photometric and spectroscopic observations, our goal is to use EW values for LAEs and EW limits for candidates with non-detections to perform the neutral fraction inference. The EW values and limits for the $z$ $\sim$ 7.6 sample are computed by \citet{hoag_constraining_2019} using the methods described therein. For the $z$ $\sim 5.5-7$ galaxies, the EW values of the 36 Ly$\alpha$ detections are calculated by \citet{fuller_spectroscopically_2020}. Here we determine limits on EW for the non-detections using a method comparable to that used by \citet{fuller_spectroscopically_2020} for the detections. 

Equivalent width is a relative value which does not require knowledge of a magnification value, nor flux calibration if there is simultaneous detection of continuum. However, in the case of faint galaxies at high redshift, we typically do not detect continuum. Instead, we must first put the error spectrum on an absolute scale, then use it in conjuction with photometry to calculate an EW limit. In the \citet{fuller_spectroscopically_2020} sample, there are no sources with detected continuum (see also \citealp{lemaux_size_2020}), so spectrophotometric calibration is necessary for EW calculations for all candidates in the samples. This calibration is therefore determined as follows.

We calculate flux limits for non-LAEs in the DEIMOS samples by calibrating with a bright source with continuum on the same mask to establish an absolute scale, then determine expected slit losses for both the bright object and LBG candidate, based on simulations described in \citet{lemaux_serendipitous_2009}. Then we apply these corrections to the noise spectrum, and the flux limit for a given observation, $f_{\textrm{lim}}$, is determined following a method effectively identical to that of \citet{hoag_constraining_2019}. We find the 1$\sigma$ Ly$\alpha$ rest-frame EW spectra via:

\begin{equation}
\textrm{EW}_{lim} = \frac{f_{\textrm{lim}}(\lambda)}{f_{\textrm{cont}}(1+z)}
\end{equation}

We determine the continuum flux density, $f_{\textrm{cont}}$, defined as
\begin{equation}\label{eq:3}
f_{\textrm{cont}} = 10^{-0.4(m_\textsc{AB}+48.6)}c/\lambda^2
\end{equation}
from photometry. In this definition, $m_\textsc{ab}$ is the apparent magnitude in the band used to determine continuum flux, typically the F105W band for the $z$ $\sim$ 5.5-7 sources, corresponding to an average rest-frame wavelength of $\lambda \sim$ 1500\AA. If there is no F105W data available, F125W (rest-frame $\lambda \sim$ 1800\AA) is used, and, in a few cases, where there is neither an F150W nor F125W magnitude, F140W (rest-frame $\lambda \sim$ 2000\AA) is used. For non-detections, the entire EW spectrum over the DEIMOS window is used in the inference to account for wavelength-dependent sensitivities. We considered using the method of estimating EW by extrapolating the continuum flux to the Ly$\alpha$ wavelength and determining longer wavelength continuum values via the galaxy's $\beta$ slope, but for this analysis, our uncertainties on magnitudes, and therefore $\beta$, are too high to attempt this type of precision correction. Having a large pool of non-detections and their upper limits at various UV magnitudes is helpful to constrain the EW distribution which is used in our neutral fraction inference, as detailed in \citet{mason_universe_2018} and \citet{hoag_constraining_2019}. In Figure~\ref{fig:limit}, we show the 1$\sigma$ flux and EW limits for all LBG candidates without spectroscopic detection from both DEIMOS and MOSFIRE.

\section{Analysis}

\subsection{Comparison of Samples}

\begin{figure*}
\includegraphics[width=18cm]{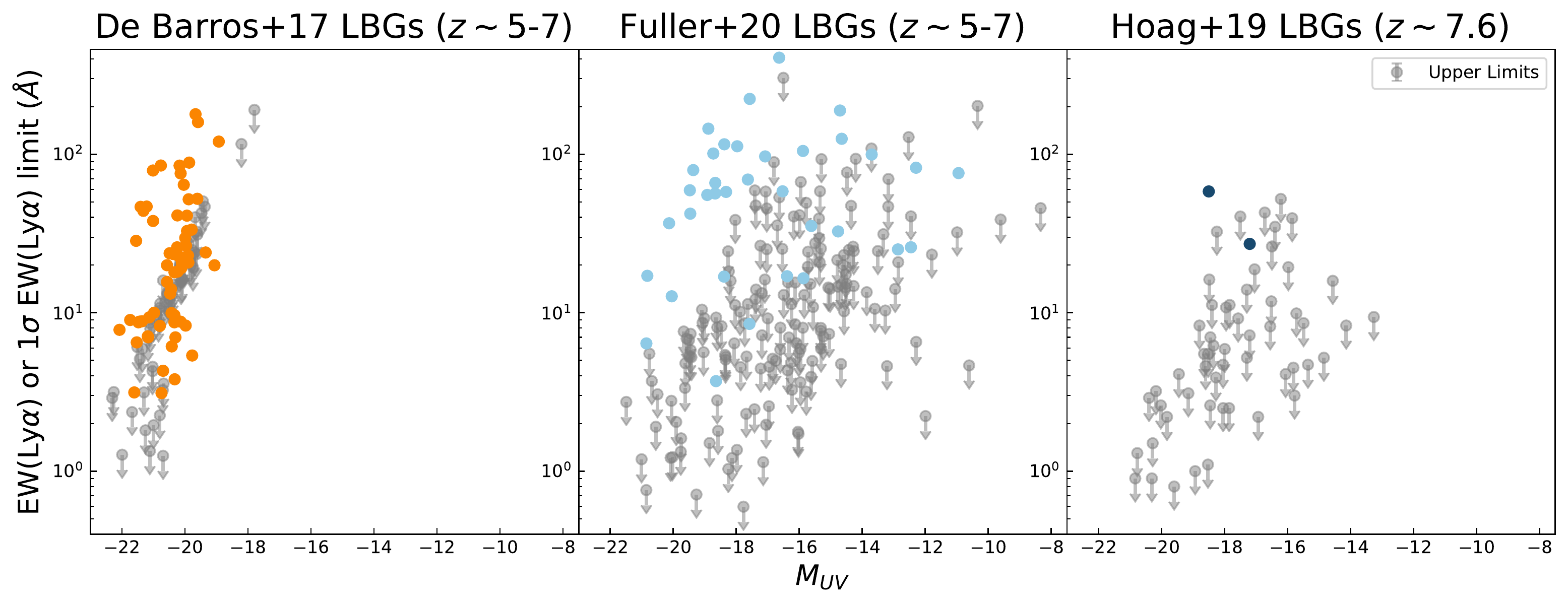}
\caption{Ly$\alpha$ EW value (LAEs, colored dots) or 1$\sigma$ upper limits (nondetections, grey arrows) of the full samples of LBG candidates from \citet{de_barros_vltfors2_2017} (left), \citet{fuller_spectroscopically_2020} (center), and \citet{hoag_constraining_2019} (right). The $z \sim 5-7$ LBG candidates from \citet{fuller_spectroscopically_2020} comprise the largest sample of galaxies at this redshift over these faint UV luminosities. We achieve comparable EW limits to those of \citet{de_barros_vltfors2_2017} but for a much fainter sample, making our three samples ($z \sim 6.0$ and $z \sim 6.7$ from \citealp{fuller_spectroscopically_2020} and $z \sim 7.6$ from \citealp{hoag_constraining_2019} comparable in $M_\textsc{uv}$.}
\label{fig:diagnostic}%
\end{figure*}

For the most accurate results in our neutral fraction inference, a good model for intrinsic Ly$\alpha$ emission after escape from the ISM is necessary. Since we compare the Ly$\alpha$ EW distribution at $z \sim$ 6.0 to that at higher redshifts, it is important to create a model for intrinsic Ly$\alpha$ emission based on a reference sample that is similar to those at high-$z$, particularly with comparable luminosities. \citet{hoag_constraining_2019} performed this $\overline{x}_\textsc{hi}$ inference based on a sample of galaxies at $z$ $\sim$ 7.6 with a bright sample of $z$ $\sim$ 6.0 LBG candidates from \citet{de_barros_vltfors2_2017} serving as the reference. These 127 galaxies are not lensed and have a median luminosity of 0.6$L^*$. We expand upon the previous analysis by including lensed galaxies from \citet{fuller_spectroscopically_2020} which represent a more typical, faint population, and have a median luminosity of 0.07$L^*$, approximately an order of magnitude fainter than the \citet{de_barros_vltfors2_2017} sample. This sample provides a similar population of galaxies to the $z$ $\sim$ 6.7 and $z$ $\sim$ 7.6 samples, that have median luminosities of 0.08$L^*$ and 0.1$L^*$, respectively. We show EW values and upper limits as a function of absolute magnitude for all three datasets in Figure~\ref{fig:diagnostic}, highlighting the depth and faintness of our new sample. The incorporation of a fainter reference sample of $z$ $\sim$ 6.0 LBG candidates as well as a more thorough treatment of the error in our results increase our confidence in the final values. 

Using the same reference sample, we infer two neutral fraction values during the EoR: one at $z$ $\sim$ 6.7, and another at $z$ $\sim$ 7.6. The total sample used to model Ly$\alpha$ EW distribution at $z \sim$ 6.0 is comprised of 83 galaxy candidates and 98 confirmed LAEs from \citet{fuller_spectroscopically_2020} and \citet{de_barros_vltfors2_2017}. We do not use the entire \citet{fuller_spectroscopically_2020} sample of 198 LBG candidates in this analysis, as we use a stricter $P(z)$ cut, $\int_{z_{\textrm{low}}}^{z_{\textrm{high}}} P(z) dz$ > 20$\%$ for the range $5.5<z<6.5$. 

In this work, we attribute any differences between the observed Ly$\alpha$ EW distribution of the reference sample and the intrinsic distribution at higher redshifts to IGM attenuation. The difference between $z$ = 6.0 and $z$ = 7.6 represents a roughly 250Myr difference in cosmic time, and the time between $z$ = 6.0 and $z$ = 6.7 is less than 200Myr. Studies indicate that the ISM does not evolve much on these timescales \citep[e.g.][]{wong_timescale_2009}, and the changes would likely be trending toward higher Ly$\alpha$ EW at increasing redshift due to a lack of dust and low metallicity \citep{hayes_redshift_2011}, resulting in our inferred neutral fraction to be a lower limit. 

To check if our reference sample has similar properties to those at higher redshifts, we compare galaxy properties which give insight into ISM conditions: the UV $\beta$ slope of the spectrum redward of Ly$\alpha$ flux and the absolute magnitude, $M_\textsc{uv}$. There is evidence that intrinsic Ly$\alpha$ strength (i.e. after escaping the ISM) is correlated with these galaxy properties \citep{oyarzun_how_2016, oyarzun_comprehensive_2017, verhamme_3d_2006, reddy_hduv_2018}. We perform statistical tests on our galaxy populations using these two properties to see if there is significant evidence of the two samples coming from the same parent distribution, as explained next. The more similar the characteristics of the two galaxy populations at different redshifts, the more confident we can be in the results of our inference, as the ISM properties are likely to not differ much. 

\subsubsection{UV Beta Slopes}

The UV continuum slope of a galaxy's spectrum, or its $\beta$ slope, characterizes its flux redward of Ly$\alpha$ emission with the relation $f_{\lambda} \propto \lambda^{\beta}$. As mentioned above, $\beta$ can give insight into a galaxy's physical properties such as star formation rate, metallicity, dust content, and age \citep[e.g.][]{buat_goods-_2012, yamanaka_uv_2019, calabro_vandels_2021}. To compare the galaxies in our three different redshift ranges, the $\beta$ slopes are computed for each LBG candidate which has the requisite data. To determine a $\beta$ slope, at least two magnitude measurements in filters redward of the expected Ly$\alpha$ emission are required. All $\beta$ slopes are calculated using linear regression fitting of the magnitude values and associated errors from Scikit-Learn \citep{scikit-learn}. 

For the sample of $z$ $\sim$ 7.6 galaxies, the available bands from the \emph{HST} Wide Field Camera 3 (WFC3/IR, \citealp{kimble_wide_2008}) for calculating $\beta$ are F125W, F140W, and F160W, as Ly$\alpha$ would fall, when present, within the F105W  band. Of the 68 candidates in this sample, 58 have magnitude values in all three bands. Two galaxies have data for only F125W and F160W filters, and we determine the slope using those two points. For the remaining eight galaxies, we do not have sufficient data to measure a $\beta$ slope. For the 198 LBG candidates at $z$ $\sim$ 6, all four of these \emph{HST} WFC3 filters can be, in principle, used in determination of a $\beta$ slope. For the 36 LAEs in the sample, 8 did not have sufficient photometric data to measure a $\beta$ slope, and of the remaining 28, 26 of them had magnitudes in all four filters, while 2 had magnitudes in just two \emph{HST} filters. Out of 162 galaxies without a Ly$\alpha$ detection, 154 had enough data to determine $\beta$, 2 based off of two filters, 17 based off of three filters, and the remaining 135 calculated from all 4 filters. We note that $\beta$ slope calculations can be influenced by the detection band (F160W for our samples) due to bias from photometric scatter. However, as all our samples use the same band for detection, our $\beta$ values would all have the same bias, so differentially there is no cause for concern. 

Rather than use a sharp cutoff on integrated $P(z)$ when performing statistical tests on galaxy properties between the reference and high-$z$ samples, we adopt a Monte Carlo (MC) method to check the similarity of two samples of candidates. For determining the similarity of $\beta$ slope distributions, we draw numbers from a random uniform sample between 0 and 1 and check if this number is less than the percentage of $P(z)$ integrated over the desired redshift range ($5.5 < z < 6.5$ for the $z \sim$ 6.0 sample, $6.5 < z < 6.9$ for the $z \sim$ 6.7 sample, and $7 < z < 8.2$ for the $z \sim$ 7.6 sample) for a given MC iteration. If it is, we keep that galaxy's beta slope in the sample. We then perform a two-sample Kolmogorov-Smirnov (KS) test between the $\beta$ slopes of the remaining $z$ $\sim$ 6.0 and $z$ $\sim$ 6.7 or $\sim$ 7.6 galaxies, and iterate this process 100 times. The error distribution on $\beta$ is also sampled over by taking the error from the linear regression fit determining $\beta$, multiplying it by a sampled value from a normalized Gaussian distribution, and adding that value to the original $\beta$. On every run of this analysis, there are no iterations out of 100 for which we reject the null hypothesis that the two samples are drawn from the same parent distribution at the 3$\sigma$ level for the $z \sim$ 6.0 and $z$ $\sim$ 6.7 samples, and $\sim$ 3 rejections for the $z \sim$ 6.0 and $z$ $\sim$ 7.6 samples. This rejection is based on the output p-value from the KS test, defined as the probability of obtaining test results at least as extreme as the results actually observed under the null hypothesis of identical parent populations. The threshold for rejection is a p-value of less than 0.005. This is true for both the original $\beta$ values and those modulated by the errors. We conclude that there is no significant evidence to reject the hypothesis that either pair of LBG candidate samples used in our inference are drawn from the same parent distribution. 

\begin{figure}
\includegraphics[width=9.5cm]{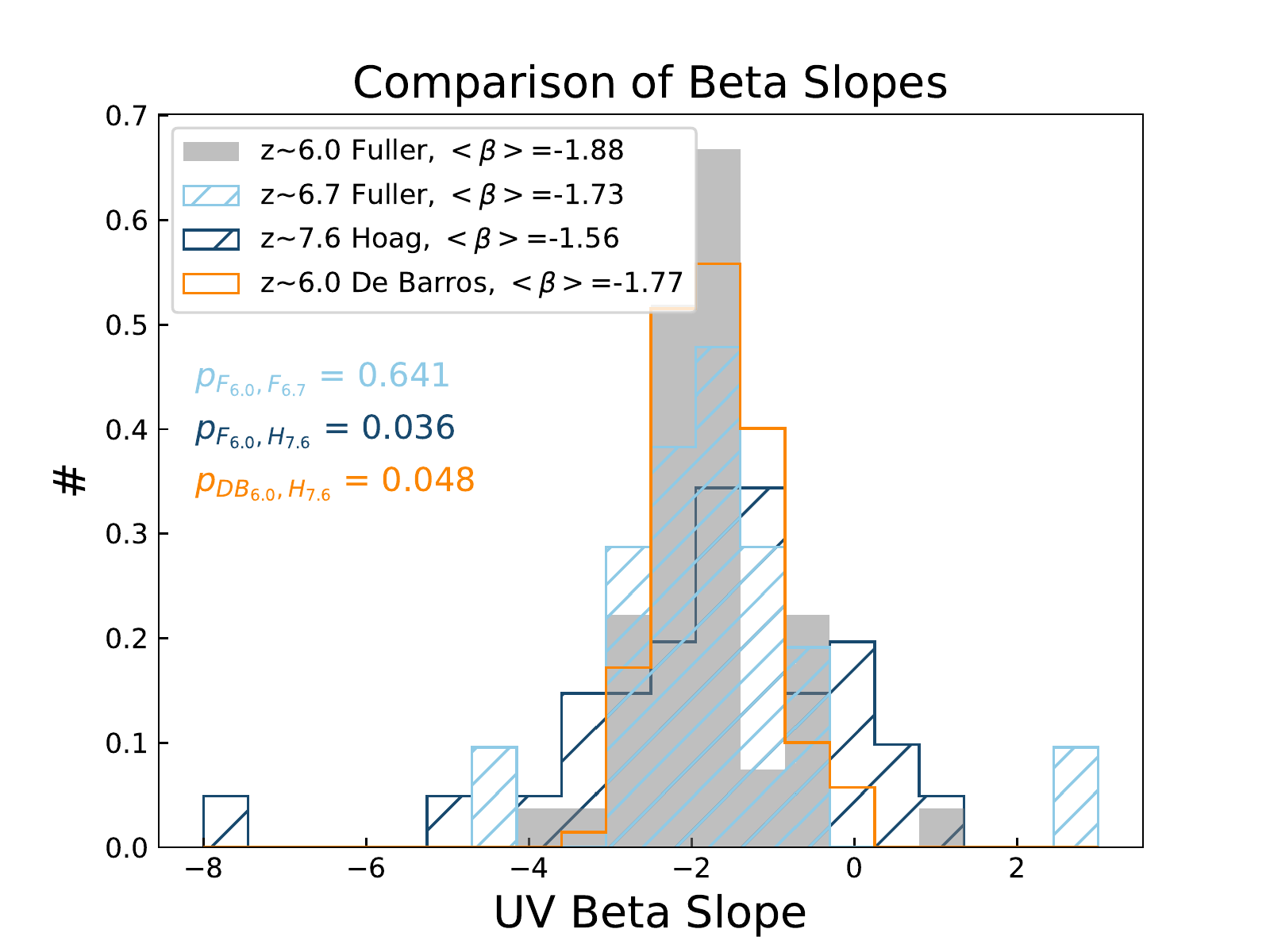}
\caption{A comparison of UV $\beta$ slope values of the three samples used in this work as well as the $z$ $\sim$ 6.0 sample used in the 2019 analysis. This plot and the corresponding p-values only include candidates with $P(z)$ > 20$\%$ in the desired redshift range. The p-values from a KS test between the $z$ $\sim$ 7.6 and each of the $z$ $\sim$ 6.0, as well as that from between the \citet{fuller_spectroscopically_2020} 6.0 and $z$ $\sim$ 6.7 samples can be seen beneath the legend. For all three pairs of samples, there is not significant evidence to reject the null hypothesis that the distribution comes from the same parent distribution.}
\label{fig:betaslope}%
\end{figure}

For visualization purposes, we plot the distribution of $\beta$ slopes for galaxies with at least 20$\%$ of their $P(z)$ between the desired redshift range for each sample in Figure~\ref{fig:betaslope}. We also include the $\beta$ slopes from the \citet{de_barros_vltfors2_2017} sample which was used as the $z$ $\sim$ 6.0 baseline for the inference in \citet{hoag_constraining_2019}. The entire sample from \citet{de_barros_vltfors2_2017} is included as we do not have $P(z)$ information for this data. The p-values displayed on the left of the figure are from KS tests between the $z \sim$ 6.0 reference sample and each of the high-$z$ samples, as well as one between the $z$ $\sim$ 7.6 sample and the \citet{de_barros_vltfors2_2017} $z \sim$ 6.0 sample. In all three cases, there is not significant evidence to reject the null hypothesis at the 3$\sigma$ level. We note that the median $\beta$ values are consistent with \citet{bouwens_uv-continuum_2014} results, as they find that fainter galaxies at a fixed redshift typically have bluer, or more negative, $\beta$ slopes. While they also find slight reddening with cosmic time at fixed $M_\textsc{uv}$, it is not significant, especially the change between $z \sim$ 7.6 and $z \sim$ 6.0. 

\subsubsection{$M_\textsc{uv}$}
We perform a similar check on the distribution of absolute magnitude values for the galaxy candidate samples. For each galaxy, $M_\textsc{uv}$ is calculated via
\begin{multline}
       M_\textsc{uv} \approx M_\textsc{FUV} = m_{F160W} + 2.5\log_{10}(\mu) - \\
         5(\log_{10}(d_{L})-1) + 2.5\log_{10}(1+z) +0.12
\end{multline}
where $m_{F160W}$ is the magnitude in the F160W band, $\mu$ is the median magnification value, $d_L$ is luminosity distance in parsecs, and $z$ is the peak redshift of the $P(z)$ distribution, and 0.12 is a K-correction to account for the change in magnitude going from the GALEX NUV to FUV band (see \citealp{fuller_spectroscopically_2020} for details). The calculation of absolute magnitude for a galaxy depends on its redshift, both directly and through $\mu$ and the K-correction. However, the effects of changes to $\mu$ and the K-correction based on redshift uncertainties within the desired redshift range for each sample is much smaller than the rest of the uncertainties on $M_\textsc{uv}$, hence we assume these to be negligible. 

To account for uncertainties in photometric redshift of the non-emitter sample, we sample over the $P(z)$ distribution. An MC approach is utilized again, sampling over the $P(z)$ of each galaxy; if the randomly sampled redshift from the $P(z)$ distribution fits within the desired range, we keep it for that iteration, compute the $M_\textsc{uv}$ based on that value, and perform a KS test between the $M_\textsc{uv}$ distributions of the remaining galaxies from the reference $z$ $\sim$ 6.0 sample and either $z$ $\sim$ 6.7 or $z$ $\sim$ 7.6 galaxies. Once again, we also do a test with modulated errors in the same vein as with the $\beta$ slopes: we multiply the $M_\textsc{uv}$ error, with a minimum value of 0.3 magnitudes, by a randomly sampled number from a Gaussian distribution, add that number to the original $M_\textsc{uv}$ value, and perform another KS test between the two modulated error samples. This process is then iterated 100 times. For each 100 iterations of the test, there are typically about 5-10 3$\sigma$ rejections of the null hypothesis for the test between the $z$ $\sim$ 6.0 and 7.6 samples, and none for the $z$ $\sim$ 6.0 and 6.7 galaxy candidates. Based on the analysis, there is little significant evidence to suggest the populations are from different distributions. 

\begin{figure}
\includegraphics[width=9.5cm]{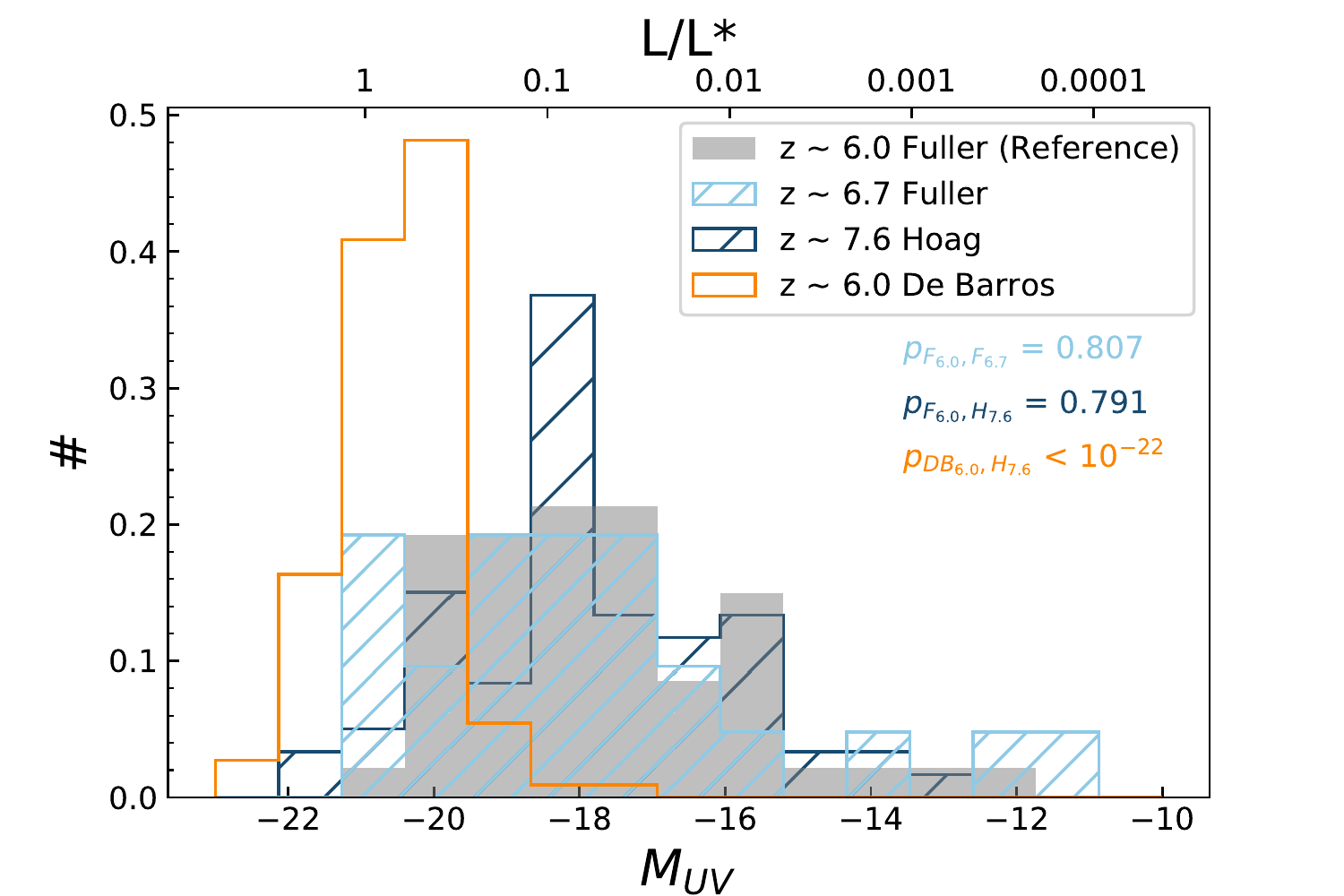}
\caption{A comparison of $M_\textsc{uv}$ values of the three samples used in this work: shaded gray ($z \sim 6.0$), hatched light blue ($ \sim$ 6.7), and dark blue ($z \sim$ 7.6) as well as the $z \sim$ 6.0 sample used in the 2019 analysis (orange). As with the $\beta$ slopes, this plot and the corresponding p-values only include candidates with greater than 20$\%$ of the $P(z)$ within the desired redshift range. The upper x-axis shows $L/L^*$, where $L^*$ is the characteristic luminosity of galaxies at $z$ = 6-8. The p-values from a KS test between the $z$ $\sim$ 6.0 and $z$ $\sim$ 6.7 samples, in addition to that between the $z$ $\sim$ 7.6 and each of the $z$ $\sim$ 6.0 samples can be seen beneath the legend. There is significant evidence that the \citet{de_barros_vltfors2_2017} and \citet{hoag_constraining_2019} are not from the same parent distribution. However, there is no indication that the latter and the \citet{fuller_spectroscopically_2020} sample are not from the same distribution.}
\label{fig:muv}%
\end{figure}

Our reference faint sample is a clear improvement from the bright \citet{de_barros_vltfors2_2017} sample for comparison to the $z\sim$ 7.6 sample, specifically in terms of absolute magnitudes. The KS test shows that the \citet{de_barros_vltfors2_2017} sample and the $z ~\sim$ 6.7 and $z~\sim$ 7.6 samples are not from the same parent distributions at high confidence. This is shown in Figure~\ref{fig:muv}, displaying the distribution of $M_\textsc{uv}$ values for all four samples, once again with a $P(z)$ cutoff of $>$ 20$\%$ in the redshift range. It is clear that the \citet{fuller_spectroscopically_2020} sample provides a much more comparable population to that of \citet{hoag_constraining_2019}, as is also evidenced by the KS test results.  
 
\color{black}

\subsection{Neutral Fraction Inference} \label{sec:nfi}

\begin{figure*}
\includegraphics[width=18cm]{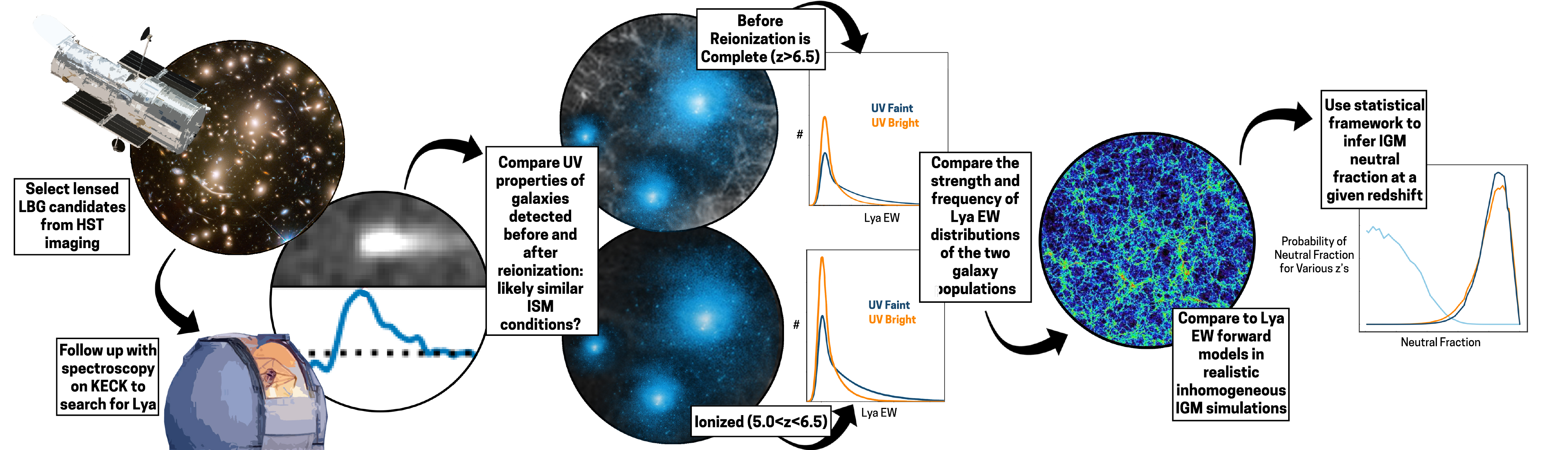}
\caption{An overview of the scientific process used in this analysis. We start by observing the fields of massive galaxy clusters which lens high-$z$ background galaxies. After selecting LBG candidates from these fields, we target them spectroscopically with DEIMOS and MOSFIRE. We then compare the ISM conditions of the LBG populations at different redshifts (section 3.1). Using the differences in Ly$\alpha$ EW distribution during and after reionization, coupled with inhomogeneous IGM simulations \citep{mesinger_evolution_2016} (image from 21cmFAST \citealp{mesinger_21cmfast_2011}), we use a Bayesian framework \citep{mason_universe_2018, mason_inferences_2019} to infer the neutral fraction of hydrogen at various EoR redshifts.}
\label{fig:inference}
\end{figure*}

In order to infer the neutral hydrogen fraction, we use the framework described by \citet{mason_universe_2018, mason_inferences_2019} and refer the reader to these papers for a full description of the methods. A comprehensive discussion of recent improvements made to the inference procedure will be provided by Mason et al, 2021 (in prep). We provide a brief overview but note that the actual inference formalism is identical to that described by \citep{mason_inferences_2019}, aside from the improvements mentioned here. In Figure~\ref{fig:inference}, we provide a schematic overview of the entire process, from LBG candidate selection to the final inference of $\overline{x}_\textsc{hi}$. Using the Ly$\alpha$ EW and $M_\textsc{uv}$ distributions of our three samples of galaxies, we determine the posterior distribution of the global IGM neutral hydrogen fraction, $\overline{x}_\textsc{hi}$ at each of the reionization era redshifts, $z \sim 6.7$ and $z\sim 7.6$. 

Our analysis uses forward models of the Ly$\alpha$ EW distribution as a function of the average IGM neutral hydrogen fraction $\overline{x}_\textsc{hi}$ and galaxies' UV luminosity to infer the evolution of $\overline{x}_\textsc{hi}$ as a function of redshift. These forward models are created using reionization simulations from the Evolution of 21cm Structure (EoS) semi-numerical simulations \citep{mesinger_evolution_2016}. These simulations generate cubes of the IGM ionization state and dark matter halos during Cosmic Dawn and the EoR using the excursion set principle and include inhomogeneous recombinations \citep{sobacchi_inhomogeneous_2014}. We populate the simulated dark matter halos with $M_\textsc{uv}$ values, Ly$\alpha$ line profiles, and EWs based on empirical models derived from our EW values and upper limits (for non-detections) in $z \sim 6.0$ galaxies (Mason et al, 2021, in prep). $M_\textsc{uv}$ values are assigned based on the relationship between halo mass and $M_\textsc{uv}$ described by \citet{mason_galaxy_2015}. 

We then calculate the observed Ly$\alpha$ EW values over multiple lines of sight after transmission through the IGM. This then provides the likelihood for our observations of each galaxy $p(W \,|\, \overline{x}_\textsc{hi}, m, \mu, z_g)$ where $W$ is the Ly$\alpha$ EW, $m$ is the apparent magnitude in the F160W band, $\mu$ is the gravitational lensing magnification, and $z_g$ is each galaxy's redshift.

To account for the unknown redshift of our non-detections we use the likelihood of observing a 1D flux density spectrum  $\{f\} = f(\lambda_i)$ for an individual galaxy (where $i$ is the wavelength pixel index), given our model where the true EW is drawn from the conditional probability distribution $p(W \,|\, \overline{x}_\textsc{hi}, m, \mu, z_g)$:
\begin{equation}
\label{eqn:inference_linelike} \begin{split}
p&(\{f\} \,|\, \overline{x}_\textsc{hi}, m, \mu, z_g, \mathrm{FWHM}) = \\
&\prod_i^N \int_0^\infty dW \, \biggl[\frac{1}{\sqrt{2\pi}\sigma_i}e^{-\frac{1}{2} \left(\frac{f_i - f_\mathrm{mod}(\lambda_i, W, m, z_d, \mathrm{FWHM})}{\sigma_i}\right)^2} \\
      	& \quad\quad\quad\quad\quad\quad \times p(W \,|\, \overline{x}_\textsc{hi}, m, \mu, z_g) \biggr] 
\end{split}
\end{equation} 
where $\sigma_i$ is the uncertainty in flux density at wavelength pixel $i$ and there are a total of $N$ wavelength pixels in the spectrum. This product of likelihoods over the wavelength range of the spectrum accounts for the wavelength sensitivity of our observations, i.e., high noise regions are weighted lower than low noise regions.

The posterior distribution for $\overline{x}_\textsc{hi}$ is obtained using Bayes' Theorem and after marginalizing over redshift $z_g$ and FWHM for each galaxy. We use a uniform prior on $\overline{x}_\textsc{hi}$ between 0 and 1, $p(\overline{x}_\textsc{hi})$, a log-normal prior on FWHM with mean depending on $M_\textsc{uv}$ as derived by empirical relations and 0.3 dex width \citep[see Appendix C by][]{mason_inferences_2019} and the photometric redshift distribution for the prior $p(z_g)$. Assuming all observations are independent, the final posterior is the product of the normalised posteriors \citep[Equation 7 by][]{mason_inferences_2019} for each object.

\subsubsection{Updates to the Inference Method}
A detailed account of all the improvements made to the inference framework since previous analyses \citep{mason_universe_2018, mason_inferences_2019, hoag_constraining_2019} will be presented by Mason et al., 2021 (in prep). The major changes are to the determination of the empirical intrinsic (before IGM absorption) Ly$\alpha$ EW distribution, but there are other smaller changes primarily concerned with accounting for all possible sources of error in the inference. Here, we give a brief overview of the improvements relevant to this analysis. 

The inclusion of a faint LBG sample, presented here, in the intrinsic Ly$\alpha$ EW model provides a more robust comparison for observations of lensed galaxies at higher redshifts \citep[e.g.][]{mason_inferences_2019,hoag_constraining_2019}. We perform meticulous checks on the data we use to determine the empirical EW distribution. The sample of $z \sim 6.0$ candidates are those from \citet{fuller_spectroscopically_2020} between $5.5<z<6.5$. We use two ways to determine which non-emitter candidates are included in this sample: selecting galaxies whose peak $P(z)$ value is between $5.5<z<6.5$ and looking at the fraction of integrated $P(z)$ between those redshifts. We find no significant change in the EW distribution parameters by using either the peak $z_{\textrm{phot}}$ or candidates whose total $P(z)$ between $z$ = 5.5 and 6.5 is greater than 20$\%$. We also experiment with removing EW data from LAEs flagged as low quality detections by \citet{fuller_spectroscopically_2020} from the analysis entirely. Removing these in the determination of the EW distribution does not significantly change the fit parameters and does not affect our $\overline{x}_\textsc{hi}$ inference. Another check we perform is testing whether changes to the error on $M_\textsc{uv}$ significantly affect the inference results. Using a minimum $M_\textsc{uv}$ error value of 0.3 to account for uncertainties in redshift within the $5.5<z<6.5$ range yields no significant change in the posteriors, compared to using $M_\textsc{uv}$ errors calculated via uncertainties in the apparent magnitude, most probable redshift, and magnification. 

Our updated method also includes propagation of errors in the intrinsic Ly$\alpha$ EW distribution into the final $\overline{x}_\textsc{hi}$ posterior. The data we use to determine the EW distribution at $z\sim6.0$ have inherent uncertainties attached to them, which is now propagated in addition to the previous inclusion of uncertainties in $m_\textsc{ab}$, $\mu$, $z$, line FWHM and line-of-sight variation. The propagation of Ly$\alpha$ EW distribution errors naturally increases the confidence intervals on $\overline{x}_\textsc{hi}$: previously, \citet{hoag_constraining_2019} inferred a neutral fraction at $z$ $\sim$ 7.6 of $\overline{x}_\textsc{hi} = 0.88^{+0.05}_{-0.10}$. Running our improved analysis at $z \sim$ 7.6 with the same data (only the \citealp{de_barros_vltfors2_2017} sample as the $z \sim$ 6.0 reference sample) yields a $\overline{x}_\textsc{hi}$ of $0.83^{+0.09}_{-0.13}$. We note that the decrease in the inferred median $\overline{x}_\textsc{hi}$ value in our work, presented below, is due to a correction of an error in the intrinsic EW distribution presented by \citet{mason_universe_2018}. The inclusion of a faint reference sample reduces the confidence interval on $\overline{x}_\textsc{hi}$ compared to using the brighter reference sample alone because the intrinsic EW distribution is better constrained at low UV luminosities. This is discussed further below (Section~\ref{sec:res})

\section{Results and Discussion} \label{sec:res}

\begin{figure}
\includegraphics[width=9.5cm]{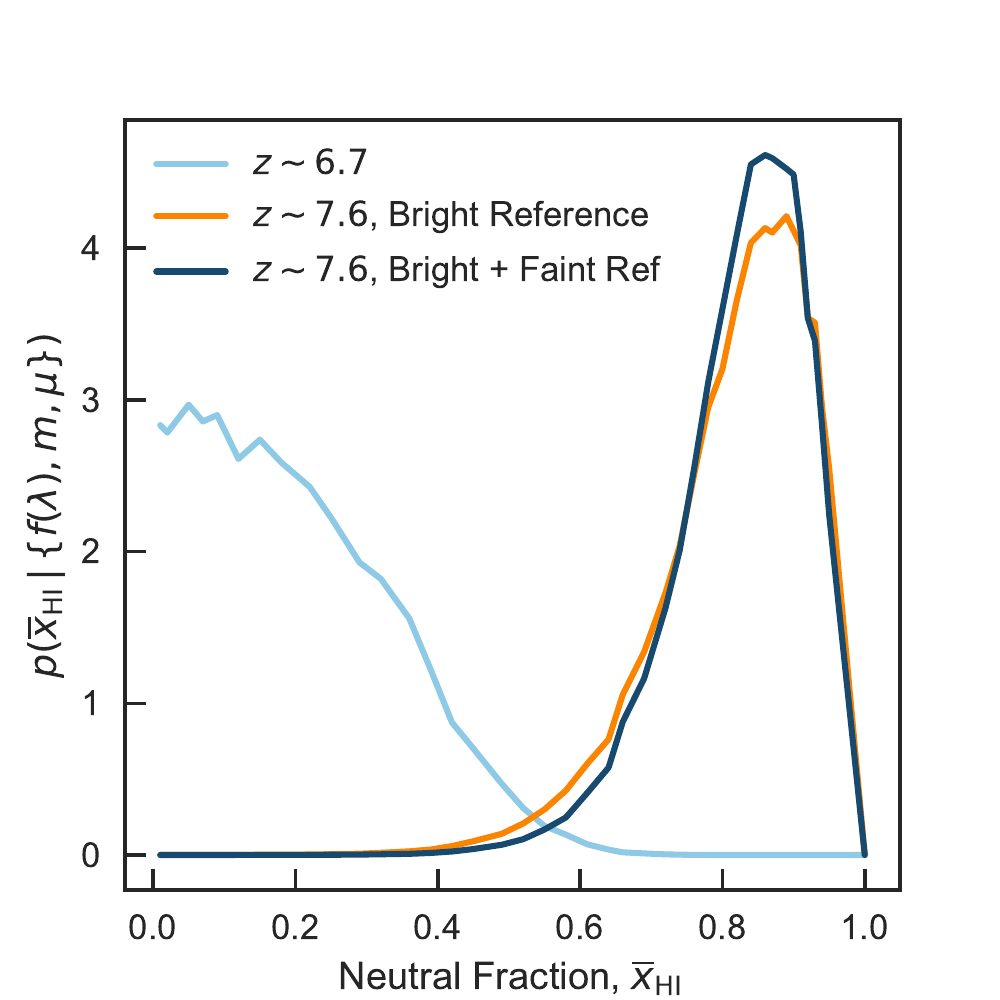}
\caption{Posterior distributions of $\overline{x}_\textsc{hi}$ for galaxies at $z \sim$ 6.7 (light blue) and at $z \sim$ 7.6 (dark blue and orange). At $z \sim$ 7.6, we show posteriors for the inferences using only the bright \citet{de_barros_vltfors2_2017} $z \sim$ 6.0 reference sample (orange) and those made using both the bright and faint \citet{fuller_spectroscopically_2020} galaxies (dark blue). By adding in a large sample of faint galaxy candidates, we reduce the uncertainty in our posterior distribution, as the curve derived from this work represents a tighter posterior.}
\label{fig:posterior}
\end{figure}

The final results of this work are an inferred upper limit on the neutral fraction at $z \sim$ 6.7 of $\overline{x}_\textsc{hi} \leq$ 0.25 within 68$\%$ uncertainty ($\leq$ 0.44 within 95$\%$ uncertainty) and a neutral fraction of $0.83^{+0.08}_{-0.11}$ at $z$ $\sim$ 7.6, with the 1$\sigma$ uncertainties derived from error propagation throughout the analysis. The final posteriors can be  seen in Figure~\ref{fig:posterior}. We include the $z \sim$ 7.6 posterior derived using both the bright \citet{de_barros_vltfors2_2017} reference sample and our faint \citet{fuller_spectroscopically_2020} $z \sim$ 6.0 sample on top of that computed using only the bright reference sample. The results using only bright galaxies in the reference sample yields a $\overline{x}_\textsc{hi}$ of $0.83^{+0.09}_{-0.13}$. With our inclusion of a faint reference sample, the inferred neutral fraction is $0.83^{+0.08}_{-0.11}$.

\begin{figure*}
\includegraphics[width=15cm]{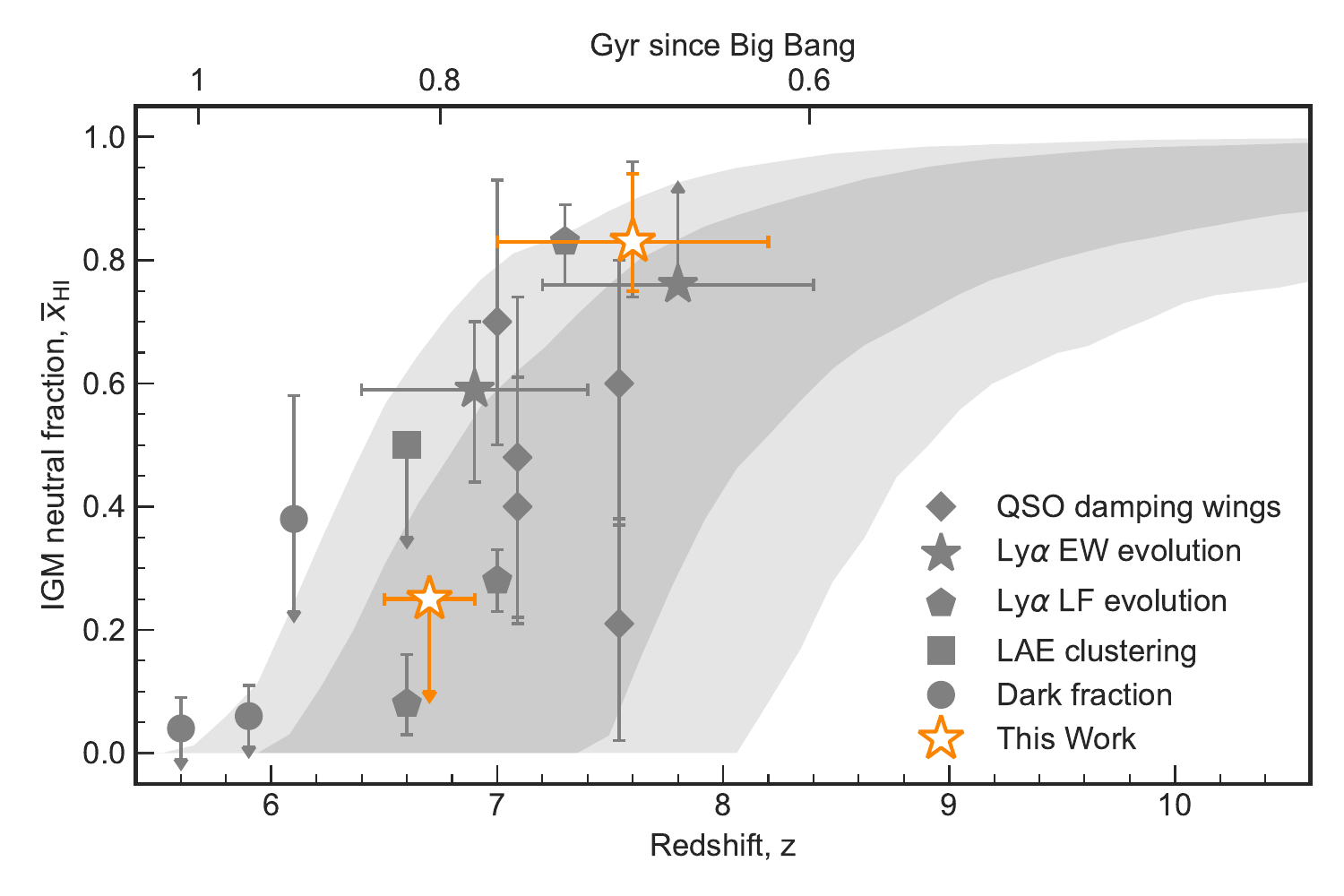}
\caption{New measurements of $\overline{x}_\textsc{hi}$ (orange stars) compared to values derived from other studies. All error bars and upper limits correspond to 1$\sigma$. The 68 and 95 percent confidence intervals on the reionization history are shaded in grey, calculated from \citet{planck_collaboration_planck_2016} CMB optical depth and dark pixel fraction constraints \citep{mason_model-independent_2019}. The grey stars represent inferred $\overline{x}_\textsc{hi}$ values using the same Bayesian analysis as this work \citep{mason_universe_2018, mason_inferences_2019, hoag_constraining_2019}. Note the point from the previous analysis \citep{hoag_constraining_2019} at the same point as this work but with larger error bars. Other data points are from a related inference method which incorporates the evolution of the Ly$\alpha$ luminosity function \citep{morales_evolution_2021}, the clustering of LAEs as squares \citep{ouchi_statistics_2010, sobacchi_clustering_2015}, Ly$\alpha$ and Ly$\beta$ forest dark pixel fraction in circles \citep{mcgreer_model-independent_2015}, and diamonds representing quasar damping wings \citep{davies_quantitative_2018, greig_constraints_2019, wang_significantly_2020}.}
\label{fig:reionhistory}
\end{figure*}

By adding in a large sample of intrinsically faint galaxies at $z \sim$ 6.0, we reduce uncertainties in the inference of $\overline{x}_\textsc{hi}$ at $z \sim$ 6.7 and $z$ $\sim$ 7.6 by better constraining the EW distribution of Ly$\alpha$ before transmission through the IGM. Our errors on the neutral fraction at $z \sim 7.6$ reflect a 14$\%$ reduction in uncertainty from the \citet{hoag_constraining_2019} analysis using the same $z \sim 7.6$ sample. Figure~\ref{fig:reionhistory} shows our two values alongside others derived from a range of methods which employ inhomogeneous models of reionization history. 

Our upper limit on the neutral fraction at $z \sim$ 6.7 is consistent within other limits placed at similar redshifts. As we get closer to the end of the EoR, it becomes more difficult to place a lower limit on $\overline{x}_\textsc{hi}$ due to intervening remaining patches of neutral hydrogen. Our upper limit is lower than that found using a group of clustered LAEs at $z \sim$ 6.6 \citep{ouchi_statistics_2010, sobacchi_clustering_2015}. By including the information from a full sample of LBG candidates, even those without detected Ly$\alpha$ emission, we are able to place tighter constraints on the neutral fraction. \citet{morales_evolution_2021} recently performed similar inferences on $\overline{x}_\textsc{hi}$ during reionization using realistic models of the Ly$\alpha$ LF with Ly$\alpha$ EW models to infer neutral fractions, shown as pentagons in Figure~\ref{fig:reionhistory}, which are consistent with our values within errors. Our $z$ $\sim$ 7.6 value is higher, though not significantly, than that found using the damping wings of a bright quasar at $z$ = 7.54 \citep{banados_800-million-solar-mass_2018}, calculated in \citet{davies_quantitative_2018}: $\overline{x}_\textsc{hi} = 0.56^{+0.21}_{-0.18}$. The higher precision reflects the use of faint galaxies over multiple lines of sight with typical ionized bubbles around them, which alleviates cosmic variance. \citet{greig_constraints_2019} independently find a $\overline{x}_\textsc{hi}$ of $0.21^{+0.17}_{-0.19}$ using this same object, due to a different intrinsic Ly$\alpha$ emission model. \citet{jung_texas_2020} calculate a $\overline{x}_\textsc{hi}$ value of $0.49^{+0.19}_{-0.19}$ at $z \sim 7.6$ using Ly$\alpha$ emission from 10 galaxies at $z>7$. We do not include this value in Figure \ref{fig:reionhistory} as the analysis does not include the effects of inhomogeneous reionization or account for sightline variance. 

Shaded in grey are the 68 and 95 percent confidence intervals on the reionization history, calculated from \citet{planck_collaboration_planck_2016} CMB optical depth and dark pixel fraction constraints \citep{mason_model-independent_2019}. Our two values are consistent within 1$\sigma$ with these constraints. The evolution of the three values from \citet{morales_evolution_2021}, represented as pentagons, also supports a late and rapid reionization scenario. The simulations introduced in \citet{kannan_introducing_2021} also agree with a fast reionization history, but favor it occurring later in cosmic time. 

Our neutral fractions at $z \sim$ 6.7 and $z \sim$ 7.6 are consistent with a rapid and late reionization. A $\overline{x}_\textsc{hi}$ of $0.83^{+0.08}_{-0.11}$ at $z \sim$ 7.6 implies a universe composed of mostly neutral hydrogen at this redshift, which quickly drops within less than 100Myrs by $z \sim$ 6.7 down to one quarter or less of IGM neutrality. It is likely that young stars within galaxies, perhaps with a contribution from quasars or lower luminosity AGN (see \citealp{grazian_lyman_2016} and references therein) were emitting bulk amounts of ionizing photons into the IGM in this time period \citep[e.g.][]{bouwens_star_2003, yan_major_2004, finkelstein_evolution_2015}. It is also possible that the contribution of bright galaxies increases with cosmic time \citep[e.g.][]{smith_thesan_2021}. 

An important note on this analysis is that we are observing low-luminosity galaxies, which are likely the ones producing the majority of ionizing photons \citep{bouwens_reionization_2015, finkelstein_evolution_2015, livermore_directly_2017}. They are also in the least biased environments on average as evidenced by clustering \citep[e.g.][]{durkalec_vimos_2018}. Other probes of reionization, such as quasars and bright LBGs, investigate denser regions surrounding high-mass halos with bright luminosities. 

Our $\overline{x}_\textsc{hi}$ value is one of several at $z$ $>$ 7 \citep[e.g.][]{davies_quantitative_2018, greig_constraints_2019, hoag_constraining_2019, jung_texas_2020, wang_significantly_2020, morales_evolution_2021}. In order to determine a detailed timeline of reionization history, observations of large samples of galaxies at high redshift are imperative. Surveys similar to the ones used in this work at $z$ $\sim$ 7-10 will become more readily available with the launch of \emph{JWST}. Using a combination of NIRCAM photometric data and spectroscopy from the NIRSPEC or NIRISS instruments, we will be able to do studies similar to this one on large samples of faint LBGs at $z$ $\geq$ 8. 

While there is no change in the recovered inferred neutral hydrogen fraction of the IGM at $z$ $\sim$ 7.6 in our updated method as compared to a previous attempt, there are clear improvements to using a galaxy population which has similar characteristics to the $z$ $\sim$ 6.7 and $z \sim$ 7.6 galaxies that allow us to better characterize several parts of this analysis. These improvements will prove even more important for future larger samples of LBGs. The consistency in $\overline{x}_\textsc{hi}$ between the previous approach and our updated analysis speaks to the robustness of the inference method. By introducing a large sample of low-luminosity galaxy candidates to the inference method, the EW distribution in the ionized universe can be more tightly constrained, leading to a smaller uncertainty in the final posterior. 

\section{Conclusions}
We have combined a lensed, intrinsically faint sample of nearly 200 LBG candidates \citep{fuller_spectroscopically_2020} with 68 low-luminosity candidates at $z$ $\sim$ 7.6 \citep{hoag_constraining_2019} for use in a Bayesian framework developed by \citet{mason_universe_2018, mason_inferences_2019} in order to constrain the global IGM neutral fraction, $\overline{x}_\textsc{hi}$, at $z$ $\sim$ 6.7 and $z$ $\sim$ 7.6. This work expands upon that done in \citet{mason_universe_2018} and \citet{hoag_constraining_2019} by adding a large, faint population of galaxies to the \citet{de_barros_vltfors2_2017} reference sample at $z \sim 6.0$, as well as folding in a more thorough analysis of the various uncertainties and assumptions inherent in this method. A summary of our conclusions is as follows:
\begin{itemize}
  \item We find no significant evidence of a difference between the faint $z \sim$ 6.0 reference sample and those of the two higher redshift samples based on $\beta$ slope and $M_\textsc{uv}$. This result comes from comparing the UV $\beta$ slope and the $M_\textsc{uv}$ distributions of the three samples using a Monte Carlo method to account for uncertainties in the $P(z)$ distributions for individual LBG candidates. 
  \item Inclusion of the \citet{fuller_spectroscopically_2020} sample of $z$ $\sim$ 6.0 galaxies yields the same neutral fraction at $z \sim$ 7.6 as the analysis using only the \citet{de_barros_vltfors2_2017} sample at $z$ $\sim$ 6.0, but with a 14$\%$ reduction in uncertainty. In this work, we infer a $\overline{x}_\textsc{hi}$ of $0.83^{+0.08}_{-0.11}$, compared to $\overline{x}_\textsc{hi} = 0.83^{+0.09}_{-0.13}$ without using the faint sample.
  \item We place an upper limit on $\overline{x}_\textsc{hi}$ at $z$ $\sim$ 6.7 of 0.25 within 68 $\%$ uncertainty and 0.44 within 95 $\%$ uncertainty. These two results are consistent with other studies at similar redshifts and implies rapid reionization. 
  \item Incorporating low-luminosity galaxies yields a higher precision on the neutral fraction at $z$ $\sim$ 7.6 than other studies at similar redshift. We are probing typical galaxies during the EoR, as faint galaxies exist in much higher numbers than their bright counterparts. Our $\overline{x}_\textsc{hi}$ reflects a globally-averaged neutral fraction derived from multiple sightlines. 
\end{itemize}

Over the next few years, we plan to take follow-up spectroscopic observations with Keck, \emph{JWST}, and ALMA, to target other nebular emission lines such as [CII] and CIII] in the LBG candidates in these datasets. Once precise redshifts are conclusively determined, the statistical power of this work will dramatically improve by  removing impurities and no longer requiring cuts on the integrated $P(z)$. With future larger samples, especially at $z$ $\geq$ 7, this method is promising for the measurement of $\overline{x}_\textsc{hi}$ over many EoR redshifts to constrain a detailed timeline of cosmic history during reionization. 

\section*{Acknowledgements}



This material is based upon work supported by the National Science Foundation under Grant No. AST-1815458 to M.B. and grant AST1810822 to T.T., and by NASA through grant NNX14AN73H to M.B. C.M. acknowledges support provided by NASA through the NASA Hubble Fellowship grant HST-HF2-51413.001-A awarded by the Space Telescope Science Institute, which is operated by the Association of Universities for Research in Astronomy, Inc., for NASA, under contract NAS5-26555, and by VILLUM FONDEN through research grant 37459. Based on spectrographic data obtained at the W.M.Keck Observatory, which is operated as a scientific partnership among the California Institute of Technology, the University of California, and the National Aeronautics and Space Administration. The Observatory was made possible by the generous financial support of the W.M. Keck Foundation. We thank the indigenous Hawaiian community for allowing us to be guests on their sacred mountain, a privilege, without which, this work would not have been possible. We are most fortunate to be able to conduct observations from this site. Also based on observations made with the NASA/ESA Hubble Space Telescope, obtained at the Space Telescope Science Institute, which is operated by the Association of Universities for Research in Astronomy, Inc., under NASA contract NAS5-26555. And based on observations made with the Spitzer Space Telescope, which is operated by the Jet Propulsion Laboratory, California Institute of Technology under a contract with NASA. Support for this work was also provided by NASA/HST grant HST-GO-14096, and through an award issued by JPL/Caltech.

This work used the following software packages: AstroPy \citep{2013A&A...558A..33A},  SExtractor \citep{1996A&AS..117..393B}, iPython \citep{perez_ipython_2007}, Matplotlib \citep{hunter_matplotlib_2007}, and NumPy \citep{van_der_walt_numpy_2011}.
\section*{Data Availability}
All characteristics for the $z \sim$ 7.6 galaxies are presented in Table 5 by \citet{hoag_constraining_2019}, and those of the LAEs from the DEIMOS sample are summarized in Table 3 by \citet{fuller_spectroscopically_2020}. Data describing the non-emitters from \citet{fuller_spectroscopically_2020} are summarized in a table uploaded to CDS.

\bibliographystyle{mnras}
\bibliography{Bolanetal} 




\appendix


\bsp	
\label{lastpage}
\end{document}